\documentstyle [12pt]{article}
\begin{document}
\title{ Scale Invariance, New Inflation and Decaying $\Lambda$-Terms 
\protect\\  } \author{E.I. Guendelman  
\\{\it Physics Department, Ben-Gurion University, Beer-Sheva
84105, Israel}}

\maketitle
\bigskip

\begin{abstract}
Realizations of scale invariance are studied in the context of
a gravitational theory where the action (in the first order formalism) is
of the form  $S = \int L_{1} \Phi d^{4}x$ +
$\int L_{2}\sqrt{-g}d^{4}x$ where $\Phi$ is a density built
out
of degrees of freedom, the "measure fields" independent of $g_{\mu\nu}$ 
and matter
fields appearing in $L_{1}$, $L_{2}$. If  $L_{1}$ contains the curvature,
scalar
potential $V(\phi)$ and kinetic term for $\phi$, $L_{2}$ another
potential
for $\phi$, $U(\phi)$,
then the true vacuum state has zero energy density, when theory is
analyzed in the conformal Einstein frame (CEF), where the equations assume
the Einstein form. Global Scale invariance is realized when $V(\phi)$ =
$f_{1}e^{\alpha\phi}$
and $U(\phi)$ = $f_{2}e^{2\alpha\phi}$. In the CEF  the scalar field
potential energy $V_{eff}(\phi)$
has in, addition to a minimum at zero, a flat region for  $\alpha\phi
\rightarrow\infty$, with
non zero vacuum energy, which is suitable for either a New Inflationary
scenario for the Early Universe or for a slowly rolling decaying
$\Lambda$-scenario for the late universe, where the smallness of the
vacuum energy
can be understood as a kind of see-saw mechanism. 
\end{abstract}

\section{Introduction}

Recent developments in cosmology have been influenced to a great
extent by the idea of inflation$^1$, which provides an attractive
scenario for solving
 some of the fundamental puzzles of the standard Big Bang model,
like the horizon and the flatness problems as well as providing a
framework for sensible calculations of primordial density perturbations.

	However, although the inflationary scenario is very attractive, it
has been recognized that a successful implementation requires some very
special restrictions on the dynamics that drive inflation. In particular,
in New Inflation$^2$, a potential with a large flat region, which then
drops
to zero (or almost zero) in order to reproduce the vacuum with almost zero
(in Planck units) cosmological constant of the present universe, is
required. It is hard to find a theory that gives a potential of this type
naturally.
 
	In addition to this, it is worthwhile pointing out that a
potential
with a very flat region, slowly approaching zero could be of use as a
model for a decaying cosmological constant being considered as a model for
the "accelerating universe",  now preferred by observations$^3$. This
is of
course, at a totally different scale to that of Inflation.

	Here we want to see whether such shapes of potentials can be
obtained from first principles, i.e. whether there is some fundamental
principle that produces this type of behavior for a scalar field. 

	We find indeed that this is possible and the fundamental principle
in question is none other than scale invariance. However, scale invariance
has to be discussed in a more general framework then that of the standard
Lagranian formulation of generally relativistic theories. Before going
into the question of scale invariance it is necessary therefore to discuss
first the general framework where this discussion will be set.

\section{The Non Gravitating Vacuum Energy (NGVE) Theory. Strong and Weak Formulations.}

	When formulating generally covariant Lagragian formulations of
gravitational theories, we usually consider the form
\begin{equation}
S_{1} = \int{L}\sqrt{-g} d^{4}x, g =  det g_{\mu\nu}
\end{equation}

	As it is well known, $d^{4}x$ is not a scalar but the combination
$\sqrt{-g} d^{4} x$ is a scalar. Inserting $\sqrt{-g}$,
which has the transformation properties of a
density, produces a scalar action (1), provided L is a scalar.

	One could use nevertheless other objects instead of
$\sqrt{-g}$, provided
they have the same transformation properties and achieve in this way a
different generally covariant formulation.

	For example, given 4-scalars $\varphi_{a}$ (a = 
1,2,3,4), one can construct the density
\begin{equation} 
\Phi =  \varepsilon^{\mu\nu\alpha\beta}  \varepsilon_{abcd}
\partial_{\mu} \varphi_{a} \partial_{\nu} \varphi_{b} \partial_{\alpha}
\varphi_{c} \partial_{\beta} \varphi_{d}  
\end{equation}
and consider instead of (1) the action$^4$
\begin{equation}
S_{2} =  \int L \Phi d^{4} x.	
\end{equation}
L is again some scalar, which contains the curvature (i.e. the
gravitational contribution) and a matter contribution, as is standard also
in (1).

	In the action (3) the measure carries degrees of freedom
independent of that of the metric and that of the matter fields. The most
natural and successful formulation of the theory is achieved when the
connection is also treated as an independent degree of freedom. This is
what is usually referred to as the first order formalism.

	One can notice that $\Phi$ is the total derivative of something,
for
example, one can write
\begin{equation}
\Phi = \partial_{\mu} ( \varepsilon^{\mu\nu\alpha\beta}
\varepsilon_{abcd} \varphi_{a}
  \partial_{\nu} \varphi_{b}
\partial_{\alpha}
\varphi_{c} \partial_{\beta} \varphi_{d}). 
\end{equation}

	This means that a shift of the form
\begin{equation}	
		L \rightarrow  L  +  constant	
\end{equation}
just adds the integral of a total divergence to the action (3) and it does
not affect therefore the equations of motion of the theory. The same
shift, acting on (1) produces an additional term which gives rise to a
cosmological constant.  
Since the constant part of L does not affect the equations of motion, this
theory is called the Non Gravitating Vacuum Energy (NGVE) Theory$^4$.

	One can generalize this structure and allow both geometrical
objects to enter the theory and consider
\begin{equation}
S_{3} = \int L_{1} \Phi  d^{4} x  +  \int L_{2} \sqrt{-g}d^{4}x		
\end{equation}

	Now instead of  (5),  the shift symmetry can be applied only on
$L_{1}$ 
($L_{1} \rightarrow L_{1}$ + constant). Since the structure has been
generalized, we call
this formulation the weak version of the NGVE - theory. Here $L_{1}$ and
$L_{2}$ are
$\varphi_{a}$  independent.

	There is a good reason not to consider mixing of  $\Phi$ and
$\sqrt{-g}$ , like
for example using
\begin{equation}
\frac{\Phi^{2}}{\sqrt{-g}} 
\end{equation}		 	

	This is because (6) is invariant (up to the integral of a total
divergence) under the infinite dimensional symmetry
\begin{equation}
\varphi_{a} \rightarrow \varphi_{a}  +  f_{a} (L_{1})	
\end{equation}
where $f_{a} (L_{1})$ is an arbitrary function of $L_{1}$ if $L_{1}$ and
$L_{2}$ are $\varphi_{a}$
independent. Such symmetry (up to the integral of a total divergence) is
absent if mixed terms (like (7)) are present.  Therefore (6) is considered
for the case when no dependence on the measure fields (MF) appears in
$L_{1}$ or $L_{2}$.

	In this paper  we will see that the existence of two independent
measures of integrations as in (6) allows new realizations of global scale
invariance with most interesting consequences when the results are viewed
from the point of view of cosmology.

\section{Dynamics of a Scalar Field and the Requirement of 
Scale Invariance in the Weak NGVE - Theory}

\section{The Action Principle}

	We will study now the dynamics of a scalar field $\phi$ interacting
with gravity as given by the following action
\begin{equation}
S_{\phi} =  \int L_{1} \Phi d^{4} x  +  \int L_{2} \sqrt{-g}   d^{4} x
\end{equation}
\begin{equation}
L_{1} = \frac{-1}{\kappa} R(\Gamma, g) + \frac{1}{2} g^{\mu\nu}
\partial_{\mu} \phi \partial_{\nu} \phi - V(\phi) 
\end{equation}	
\begin{equation}
L_{2} = U(\phi)
\end{equation}
\begin{equation}	
R(\Gamma,g) =  g^{\mu\nu}  R_{\mu\nu} (\Gamma) , R_{\mu\nu}
(\Gamma) = R^{\lambda}_{\mu\nu\lambda}
\end{equation}
\begin{equation}
R^{\lambda}_{\mu\nu\sigma} (\Gamma) = \Gamma^{\lambda}_
{\mu\nu,\sigma} - \Gamma^{\lambda}_{\mu\sigma,\nu} +
\Gamma^{\lambda}_{\alpha\sigma}  \Gamma^{\alpha}_{\mu\nu} -
\Gamma^{\lambda}_{\alpha\nu} \Gamma^{\alpha}_{\mu\sigma}.	 
\end{equation}

	In the variational principle $\Gamma^{\lambda}_{\mu\nu},
g_{\mu\nu}$, the measure fields scalars
$\varphi_{a}$ and the "matter" - scalar field $\phi$ are all to be treated
as independent
variables although the variational principle may result in equations that
allow us to solve some of these variables in terms of others.

\section{Global Scale Invariance}

	If we perform the global scale transformation ($\theta$ =
constant) 
\begin{equation}
g_{\mu\nu}  \rightarrow   e^{\theta}  g_{\mu\nu}	
\end{equation}
then (9) is invariant provided  $V(\phi)$ and $U(\phi)$ are of the
form  
\begin{equation}
V(\phi) = f_{1}  e^{\alpha\phi},  U(\phi) =  f_{2}
e^{2\alpha\phi}
\end{equation}
and $\varphi_{a}$ is transformed according to
\begin{equation}
\varphi_{a}   \rightarrow   \lambda_{a} \varphi_{a}  
\end{equation}
(no sum on a) which means
\begin{equation}
\Phi \rightarrow \biggl(\prod_{a} {\lambda}_{a}\biggr) \Phi \\ \equiv \lambda 
\Phi	 \end{equation}
such that
\begin{equation} 
\lambda = e^{\theta}
\end{equation}	
and	 
\begin{equation}
\phi \rightarrow \phi - \frac{\theta}{\alpha}.     	
\end{equation}

\section{The Equations of Motion}

	We will now work out the equations of motion for arbitrary choice
of $V(\phi)$ and $U(\phi)$. We study afterwards the choice (15) which
allows us to
obtain the results for the scale invariant case and also to see what
differentiates this from the choice of arbitrary $U{\phi}$ and  $V{\phi}$ 
in a very
special way.

	Let us begin by considering the equations which are obtained from
the variation of the fields that appear in the measure, i.e. the
$\varphi_{a}$
fields. We obtain then  
\begin{equation}		
A^{\mu}_{a} \partial_{\mu} L_{1} = O   	
\end{equation}
where  $A^{\mu}_{a} = \varepsilon^{\mu\nu\alpha\beta}
\varepsilon_{abcd} \partial_{\nu} \varphi_{b} \partial_{\alpha}
\varphi_{c} \partial_{\beta} \varphi_{d}$ (21). Since it is easy to
check that  $A^{\mu}_{a} \partial_{\mu} \varphi_{a^{\prime}} =
\frac{\delta aa^{\prime}}{4} \Phi$, it follows that 
det $(A^{\mu}_{a}) =\frac{4^{-4}}{4!} \Phi^{3} \neq O$ if $\Phi\neq O$.
Therefore if $\Phi\neq O$ we obtain that $\partial_{\mu} L_{1} = O$,
 or that
\begin{equation}
L_{1} = \frac{-1}{\kappa} R(\Gamma,g) + \frac{1}{2} g^{\mu\nu}
\partial_{\mu} \phi \partial_{\nu} \phi - V = M	 
\end{equation}
where M is constant.

	Let us study now the equations obtained from the variation of the
connections $\Gamma^{\lambda}_{\mu\nu}$.  We obtain then
\begin{equation}
-\Gamma^{\lambda}_{\mu\nu} -\Gamma^{\alpha}_{\beta\mu}
g^{\beta\lambda} g_{\alpha\nu}  + \delta^{\lambda}_{\nu}
\Gamma^{\alpha}_{\mu\alpha} + \delta^{\lambda}_{\mu}
g^{\alpha\beta} \Gamma^{\gamma}_{\alpha\beta}
g_{\gamma\nu}\\ - g_{\alpha\nu} \partial_{\mu} g^{\alpha\lambda}
+ \delta^{\lambda}_{\mu} g_{\alpha\nu} \partial_{\beta}
g^{\alpha\beta}
 - \delta^{\lambda}_{\nu} \frac{\Phi,_\mu}{\Phi}
+ \delta^{\lambda}_{\mu} \frac{\Phi,_           \nu}{\Phi} =  O	
\end{equation}
If we define $\Sigma^{\lambda}_{\mu\nu}$    as
$\Sigma^{\lambda}_{\mu\nu} =
\Gamma^{\lambda}_{\mu\nu} -\{^{\lambda}_{\mu\nu}\}$
where $\{^{\lambda}_{\mu\nu}\}$   is the Christoffel symbol, we
obtain for $\Sigma^{\lambda}_{\mu\nu}$ the equation 
\begin{equation}
	-  \sigma, _{\lambda} g_{\mu\nu} + \sigma, _{\mu}
g_{\nu\lambda} - g_{\nu\alpha} \Sigma^{\alpha}_{\lambda\mu}
-g_{\mu\alpha} \Sigma^{\alpha}_{\nu \lambda}
+ g_{\mu\nu} \Sigma^{\alpha}_{\lambda\alpha} +
g_{\nu\lambda} g_{\alpha\mu} g^{\beta\gamma} \Sigma^{\alpha}_{\beta\gamma}
= O 
\end{equation}		 
where  $\sigma = 1n \chi, \chi = \frac{\Phi}{\sqrt{-g}}$.
      	
	The general solution of (23) is 
\begin{equation}
\Sigma^{\alpha}_{\mu\nu} = \delta^{\alpha}_{\mu}
\lambda,_{\nu} + \frac{1}{2} (\sigma,_{\mu} \delta^{\alpha}_{\nu} -
\sigma,_{\beta} g_{\mu\nu} g^{\alpha\beta})
\end{equation}
where $\lambda$ is an arbitrary function due to the $\lambda$ - symmetry
of the
curvature${}^{(5)}$  $R^{\lambda}_{\mu\nu\alpha} (\Gamma)$,
\begin{equation}
\Gamma^{\alpha}_{\mu\nu} \rightarrow \Gamma^{\prime \alpha}_{\mu\nu}
 = \Gamma^{\alpha}_{\mu\nu} + \delta^{\alpha}_{\mu}
Z,_{\nu}
\end{equation} 
Z  being any scalar (which means $\lambda \rightarrow \lambda + Z$).
  
	If we choose the gauge $\lambda = \frac{\sigma}{2}$, we obtain
\begin{equation}
\Sigma^{\alpha}_{\mu\nu} (\sigma) = \frac{1}{2} (\delta^{\alpha}_{\mu}
\sigma,_{\nu} +
 \delta^{\alpha}_{\nu} \sigma,_{\mu} - \sigma,_{\beta}
g_{\mu\nu} g^{\alpha\beta}).
\end{equation}

	Considering now the variation with respect to $g^{\mu\nu}$, we
obtain
\begin{equation}	 	
\Phi = (\frac{-1}{\kappa} R_{\mu\nu} (\Gamma) + \frac{1}{2} \phi,_{\mu}
\phi,_{\nu}) - \frac{1}{2} \sqrt{-g} U(\phi) g_{\mu\nu} = O
\end{equation}
solving for $R = g^{\mu\nu} R_{\mu\nu} (\Gamma)$  and introducing in
(22), we obtain
\begin{equation}
M + V(\phi) - \frac{2U(\varphi)}{\chi} = O
\end{equation}
a constraint that allows us to solve for $\chi$,
\begin{equation}
\chi = \frac{2U(\phi)}{M+V(\phi)}.
\end{equation}

	To get the physical content of the theory, it is convenient to go
to the Einstein conformal frame where 
\begin{equation}
\overline{g}_{\mu\nu} = \chi g_{\mu\nu}		    
\end{equation}
and $\chi$  given by (29b). In terms of $\overline{g}_{\mu\nu}$   the non
Riemannian contribution $\Sigma^{\alpha}_{\mu\nu}$
dissappears from the equations, which can be written then in the Einstein
form ($R_{\mu\nu} (\overline{g}_{\alpha\beta})$ =  usual Ricci tensor)
\begin{equation}
R_{\mu\nu} (\overline{g}_{\alpha\beta}) - \frac{1}{2} 
\overline{g}_{\mu\nu}
R(\overline{g}_{\alpha\beta}) = \frac{\kappa}{2} T^{eff}_{\mu\nu}
(\phi)	 	
\end{equation}
where
\begin{equation}	 
T^{eff}_{\mu\nu} (\phi) = \phi_{,\mu} \phi_{,\nu} - \frac{1}{2} \overline
{g}_{\mu\nu} \phi_{,\alpha} \phi_{,\beta} \overline{g}^{\alpha\beta}
+ \overline{g}_{\mu\nu} V_{eff} (\phi)
\end{equation}
and 	
\begin{equation}
V_{eff} (\phi) = \frac{1}{4U(\phi)}  (V+M)^{2}.
\end{equation}
	
	In terms of the metric $\overline{g}^{\alpha\beta}$ , the equation
of motion of the Scalar
field $\phi$ takes the standard General - Relativity form
\begin{equation}
\frac{1}{\sqrt{-\overline{g}}} \partial_{\mu} (\overline{g}^{\mu\nu} 
\sqrt{-\overline{g}} \partial_{\nu}
\phi) + V^{\prime}_{eff} (\phi) = O.
\end{equation} 

	Notice that if  $V + M = O,  V_{eff}  =  O$ and $V^{\prime}_{eff} 
= O$ also, provided $V^{\prime}$
is finite and $U \neq O$ there. This means the zero cosmological constant
state
is achieved without any sort of fine tuning. This is the basic feature
that characterizes the NGVE - theory and allows it to solve the
cosmological constant problem$^{4}$.

	In what follows we will study (33) for the special case of global
scale invariance, which as we will see displays additional very special
features which makes it attractive in the context of cosmology.

	Notice that in terms of the variables $\phi$,
$\overline{g}_{\mu\nu}$, the "scale"
transformation becomes only a shift in the scalar field $\phi$, since
$\overline{g}_{\mu\nu}$ is
invariant (since $\chi \rightarrow \lambda^{-1} \chi$  and $g_{\mu\nu}
\rightarrow \lambda g_{\mu\nu}$)
\begin{equation}
\overline{g}_{\mu\nu} \rightarrow \overline{g}_{\mu\nu}, \phi \rightarrow
\phi - \frac{\theta}{\alpha}.
\end{equation}

\section{Analysis of the Scale - Invariant Dynamics}

	If $V(\phi) = f_{1} e^{\alpha\phi}$  and  $U(\phi) = f_{2}
e^{2\alpha\phi}$ as
required by scale
invariance (14), (16), (17), (18), (19), we obtain from (33)
\begin{equation}
	V_{eff}  = \frac{1}{4f_{2}}  (f_{1}  +  M e^{-\alpha\phi})^{2}	
\end{equation}

	Since we can always perform the transformation $\phi \rightarrow
- \phi$ we can
choose by convention $\alpha > O$. We then see that as $\phi \rightarrow
\infty, V_{eff} \rightarrow \frac{f_{1}^{2}}{4f_{2}} =$ const.
providing an infinite flat region. Also a minimum is achieved at zero
cosmological constant for the case $\frac{f_{1}}{M} < O$ at the point 
\begin{equation}
\phi_{min}  =  \frac{-1}{\alpha} ln \mid\frac{f_1}{M}\mid.  	
\end{equation}

	Finally, the second derivative of the potential  $V_{eff}$  at the
minimum is 
\begin{equation}
V^{\prime\prime}_{eff} = \frac{\alpha^2}{2f_2} \mid{f_1}\mid^{2} > O
\end{equation}
if
$f_{2} > O$,	 	
there are many interesting issues that one can raise here. The first one
is of course the fact that a realistic scalar field potential, with
massive exitations when considering the true vacuum state, is achieved in
a way consistent with the idea (although somewhat generalized) of scale
invariance.

	The second point to be raised is that there is an infinite region
of flat potential for $\phi \rightarrow \infty$, which makes this theory
an attractive
realization of the improved inflationary model$^{2}$.

	A peculiar feature of the potential (36), is that the constant M,
provided it has the correct sign, i.e. that $f_{1}/M < O$, does not affect
the
physics of the problem. This is because if we perform a shift 
\begin{equation}
\phi \rightarrow \phi + \Delta		
\end{equation}	
in the potential (36), this is equivalent to the change in the integration
constant  M
\begin{equation}
M \rightarrow M e^{-\alpha\Delta}.	
\end{equation}

	We see therefore that if we change  M in any way, without changing
the sign of M, the only effect this has is to shift the whole potential.
The physics of the potential remains unchanged, however. This is 
reminiscent of the dilatation invariance of the theory, which involves
only a shift in $\phi$  if $\overline{g}_{\mu\nu}$   is used (see eq. (35)
).

	This is very different from the situation for two generic
functions
$U(\phi)$ and 
$V(\phi)$ in (34). There, M appears in $V_{eff}$ as a true new parameter
that
generically changes the shape of the potential $V_{eff}$, i.e. it is
impossible
then to compensate the effect of M with just a shift. For example  M will
appear in the value of the second derivative of the potential at the
minimum, unlike what we see in eq. (38), where we see that
$V^{\prime\prime}_{eff}$ (min) is M
independent.

	In conclusion, the scale invariance of the original theory is
responsible for the non appearance (in the physics) of a certain scale,
that associated to M. However, masses do appear, since the coupling to two
different measures of $L_{1}$ and $L_{2}$ allow us to introduce two
independent
couplings  $f_{1}$ and $f_{2}$, a situation which is  unlike the
standard
formulation of globally scale invariant theories, where usually no stable
vacuum state exists.

	Notice that we have not considered all possible terms consistent
with global scale invariance. Additional terms in  $L_{2}$  of the form
$e^{\alpha\phi} R$ and $e^{\alpha\phi} g^{\mu\nu} \partial_{\mu}\phi
\partial_{\nu}\phi$
 are indeed consistent with the global scale invariance
(14), (16), (17), (18), (19) but they give rise to a much more complicated
theory, which will be studied in a separate publication. There it will be
shown that for slow rolling and for $\phi \rightarrow \infty$ the
basic features of the theory
are the same as what has been studied here. Let us finish this section by
comparing the appearance of the potential $V_{eff} (\phi)$, which has
privileged
some point depending on M (for example the minimum of the potential will
have to be at some specific point), although the theory has the
"translation invariance" (35), to the physics of solitons.

	In fact, this very much resembles the appearance of solitons in a
space-translation invariant theory: The soliton solution has to be
centered at some point, which of course is not determined by the theory.
The soliton of  course breaks the space translation invariance
spontaneously, just as the existence of the non trivial potential $V_{eff}
(\phi)$
breaks here spontaneously the translations in $\phi$ space, since $V_{eff}
(\phi)$ is
not a constant. 

	Notice however, that the existence for $\phi \rightarrow \infty$,
of a flat region for
$V_{eff} (\phi)$ can be nicely described as a region where the symmetry
under
translations (35) is restored.

\section{Cosmological Applications of the Model}

	Since we have an infinite region in which $V_{eff}$ as given by
(36) is
flat $(\phi \rightarrow \infty)$, we expect a slow rolling (new
inflationary) scenario to be
viable, provided the universe is started at a sufficiently large value of
the scalar field $\phi$.

       One should  point out that the model discussed here gives a
potential with two physically relevant parameters
$\frac{f_1^{2}}{4f_{2}}$ , which represents
the value of $V_{eff}$ as $\phi \rightarrow \infty$ ,  i.e. the strength
of the false vacuum at
the
flat region and  $\frac{\alpha^{2}f_	{1}^{2}}{2f_2}$ , representing the
mass of the excitations around the
true vacuum with zero cosmological constant (achieved here without fine
tuning).

	When a realistic model of reheating is considered, one has to give
the strength of the coupling of the $\phi$ field to other fields. It
remains to
be seen what region of parameter space provides us with a realistic
cosmological model.

            Furthermore, one can consider this model as suitable for the
very late universe rather than for the early universe, after we suitably
reinterpret the meaning of the scalar field  $\phi$. 

	This can provide a long lived almost constant vacuum energy for a
long period of time, which can be small if $f_{1}^{2}/4f_{2}$ is
small. Such small energy
density will eventually disappear when the universe achieves its true
vacuum state.

	Notice that a small value of $\frac{f_{1}^{2}}{f_{2}}$   can be
achieved if we let $f_{2} >> f_{1}$. In this case
$\frac{f_{1}^{2}}{f_{2}} << f_{1}$, i.e. a very small scale for the
energy
density of the universe is obtained by the existence of a very high scale
(that of $f_{2}$) the same way as a small fermion mass is obtained in the
see-saw mechanism$^{6}$ from the existence also of a large mass scale.

\section{Acknowledgement}

I would like to thank A. Davidson and A. Kaganovich for conversations on
the subjects discussed here.

\break

\end{document}